\documentclass[reprint, amsmath,amssymb, aps]{revtex4-1}

\usepackage{graphicx} 
\usepackage[utf8]{inputenc}
\usepackage{bm}
\usepackage{hyperref}
\usepackage{xcolor}
\usepackage[normalem]{ulem}
\usepackage{amsmath}
\usepackage{float}
\usepackage{algorithm}
\usepackage{algpseudocode}
\usepackage{amsmath}

\begin{document}

\title{Universal Model of Urban Street Networks}
\author{Marc Barthelemy}
\email{marc.barthelemy@ipht.fr}
\affiliation{Universit\'e Paris-Saclay, CNRS, CEA, Institut de Physique Th\'{e}orique, 91191, 
	Gif-sur-Yvette, France}
\affiliation{Centre d'Analyse et de Math\'ematique Sociales (CNRS/EHESS) 54 Avenue de Raspail, 75006 Paris, France}

\author{Geoff Boeing}
\email{boeing@usc.edu}
\affiliation{Department of Urban Planning and Spatial Analysis, Sol Price School of Public Policy, University of Southern California, 301A Lewis Hall, Los Angeles, CA 90089-0626, USA}

\date{\today}

\begin{abstract}

Analyzing 9,000 urban areas' street networks, we identify properties, including extreme betweenness centrality heterogeneity, that typical spatial network models fail to explain. Accordingly we propose a universal, parsimonious, generative model based on a two-step mechanism that begins with a spanning tree as a backbone then iteratively adds edges to match empirical degree distributions. Controlled by a single parameter representing lattice-equivalent node density, it accurately reproduces key universal properties to bridge the gap between empirical observations and generative models.

\end{abstract}

\maketitle

\paragraph*{Introduction---}Urban planners, engineers, and geographers measure and model street networks to describe or reproduce the structural characteristics that underlie urban transport, accessibility, and sustainability. These models usually incorporate geometric and topological information, and may be empirical or generative. The former usually use governmental data or open data to describe real-world networks. Various properties of these networks were studied over the past two decades \cite{Kansky:1963,Haggett:1969,Hillier:1984,Cardillo:2006,Buhl:2006,Xie:2007,Crucitti:2006,Barthelemy:2008,Lammer:2006,Barthelemy:2013,Strano:2012,Louf:2014b,Levinson:2012,kirkley2018betweenness,chen2024global,Gudmundsson:2013,Jiang:2007,Scellato:2006,Masucci:2009,Boeing:2019,Boeing:2020,Boeing:2022}, and the temporal evolution of these structures is also studied through the digitization of historical records \cite{Perret:2015,Strano:2012,Masucci:2013,Barthelemy:2013,Burghardt:2022,Taillanter:2023,barrington2015century}.

Despite this empirical research, a gap remains in developing parsimonious models that can replicate the heterogeneity of observed urban street network properties. Such models are crucial for exploring the impact of urban form on various processes such as CO$_2$ emissions, traffic congestion, and the macroscopic fundamental diagram. Previous authors proposed simple models based on local optimization processes and the mechanics of leaf pattern formation \cite{Barthelemy:2008, Barthelemy:2009,Courtat:2011}, extended in \cite{el2022urban}. Recent AI research has synthesized street network images and graphs with deep generative models \cite{hartmann2017streetgan,lin2022exploring,li2024crtgan}. Such methods usually test the visual and statistical similarity of certain street network properties such as city block area, compactness, and aspect ratio \cite{hartmann2017streetgan}. While useful, these models require extensive input data for training and potentially fail to capture unique properties that distinguish street networks from general spatial networks \cite{hartmann2017streetgan,Kuntzsch03052016,LI2024112027,Lin03052024}.

Can a simple spatial network model capture the most essential features of real-world street networks? To address this, we must first identify which features are truly essential. The primary function of street networks is to facilitate smooth traffic flow between locations, a process often governed by shortest paths. Accordingly, a key property of street networks is the distribution of shortest paths (weighted by distance), which can be characterized by betweenness centrality (BC). We propose that traffic dynamics are primarily driven by network connectivity, best described by BC-related measures. Therefore, a generative model of street networks should prioritize reproducing the BC properties observed in real-world networks.

\paragraph*{Characterizing street networks---}We thus first identify and measure key empirical street network properties. As discussed in \cite{Barthelemy2024Review}, many measures commonly used for complex networks are not applicable to spatial networks in general, and street networks in particular, due to physical constraints \cite{Lammer:2006,Mossa:2002}. Street networks are characterized by a narrow degree distribution and a high clustering coefficient \cite{barthelemy2022spatial}. Traditional indicators from transportation geography \cite{Kansky:1963} tend to be redundant, as they largely reflect variations in the average degree. A more informative characterization of street networks relies on geometric and structural measures, such as the distribution of street segment lengths, block shapes, and street angles \cite{barthelemy2022spatial}. The most fundamental quantity, the node count $N$, has been shown to grow sublinearly with population size \cite{Strano:2012,Barthelemy:2013}, with a $1\%$ increase in urban population typically resulting in a $0.95\%$ increase in the number of intersections \cite{Boeing:2022}. Furthermore, the total street length scales as $L\sim\sqrt{A_uN}$, where $A_u$ denotes the urbanized area \cite{Samaniego:2008,Barthelemy:2008}.

\paragraph*{Degree distribution and betweenness centrality---}A basic characteristic of street networks is the fraction $f_i$ of nodes with degree $i=1,2,3,4$. Specifically, $f_1$ represents the proportion of dead ends, while $f_3$ and $f_4$ correspond to 3-way and 4-way intersections, respectively. The fraction $f_2$ of degree-2 nodes is irrelevant as these have no real impact on the graph's topology (in practical applications, these nodes are sometimes kept for describing a change to street attributes in the middle of a graph edge, such as a change in lane counts, directionality restrictions, etc., which is why $f_2$ can be slightly different from zero).

\begin{figure}[bt!]
	\centering
     \includegraphics[width=0.45\textwidth]{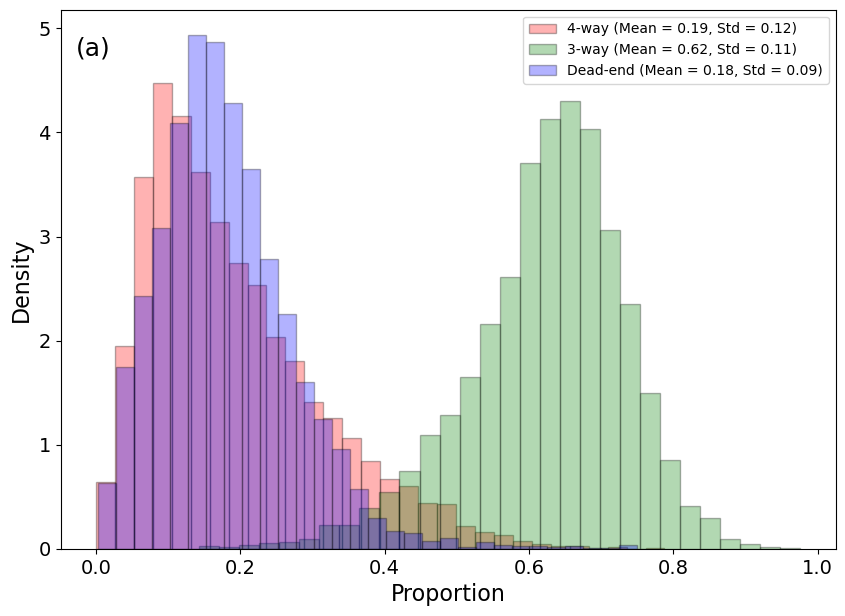}
\includegraphics[width=0.45\textwidth]{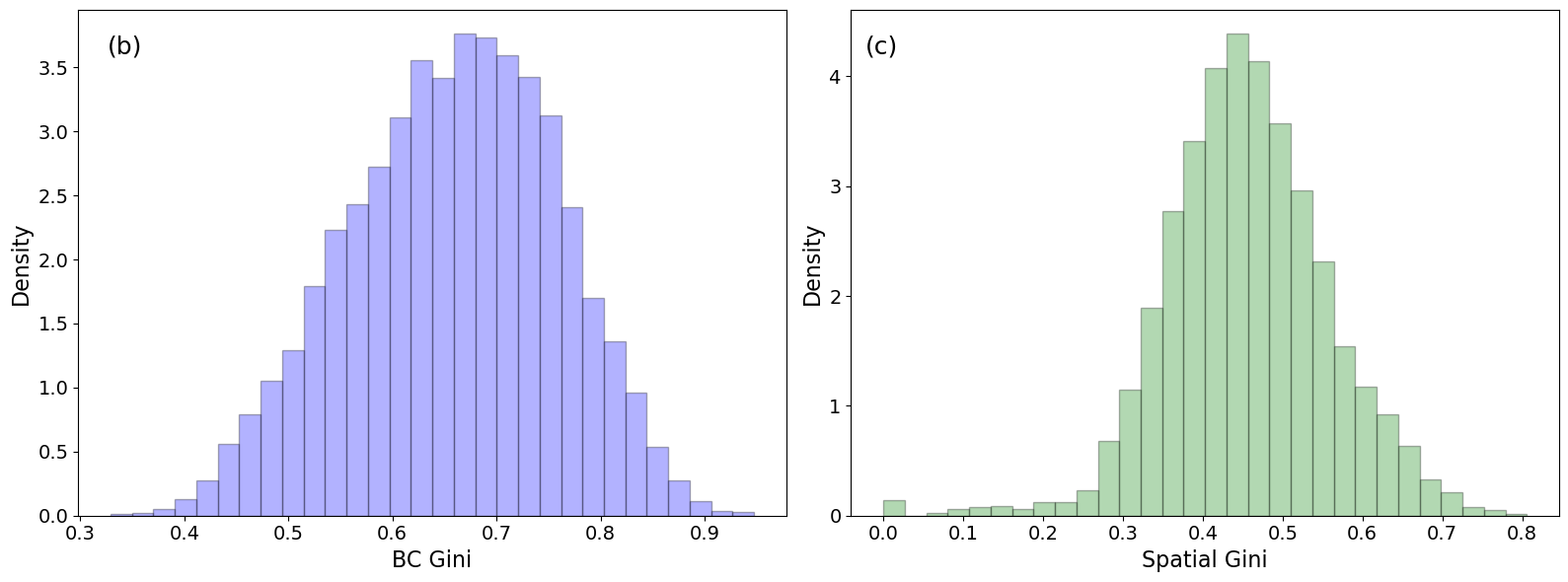}
	\caption{Empirical probability distributions of the fractions $f_i$ of nodes having a degree $i=1,3,4$ ($f_2$ not shown) for $\sim 9,000$ urban areas worldwide. The average values are $\overline{f_1}= 0.18$, $\overline{f_2}\approx 0.0$, $\overline{f_3}=0.62$, $\overline{f_4}=0.19$. (b) Probability distribution of the Gini coefficient for the BC and in (c) for the spatial Gini coefficient. }
	\label{fig:fi}
\end{figure}

An effective model should accurately reproduce the statistics of this degree distribution. We first study the empirical properties of approximately 9,000 urban areas worldwide (this dataset comes from \cite{Boeing:2022}; see also \cite{supplemental} for more details) and show that these quantities fluctuate slightly around their means (Fig.~\ref{fig:fi}). The degree distribution however does not fix the topology and many different networks are possible. An important characterization of flows on networks is given by the BC, defined for edges as \cite{Freeman:1977}
\begin{align}
    g(e) = \frac{1}{{\cal N}} \sum_{s \neq t} \frac{\sigma_{s,t}(e)}{\sigma_{s,t}},
\end{align}
where $\sigma_{s,t}$ denotes the number of shortest paths from node $s$ to node $t$, and $\sigma_{s,t}(e)$ represents the number of those paths that pass through edge $e$. 

We use shortest geometric paths that minimize total distance. Alternative criteria---such as travel time---could be considered by assigning velocities to edges. When velocities are constant, minimizing distance is equivalent to minimizing travel time. Differences arise only when velocities vary across links, as is typical in motorized transport with traffic. However, in many cases (e.g., walking), travel speeds are approximately uniform. Incorporating speed limits would constrain the model to vehicular transport and introduce additional parameters, reducing its generality and parsimony. For these reasons, we leave such extensions for future work.

Edges with high BC represent those traversed by many shortest paths, effectively functioning as main roads within the network. This quantity has been extensively studied \cite{Wu:2006,Strano:2012,kirkley2018betweenness,barthelemy2022spatial}, and a key finding is that street networks can often be decomposed into a `backbone'---typically represented by a minimum spanning tree (MST)---characterized by high BC, while the remainder of the network exhibits significantly lower BC. Edge BC correlates with age \cite{Strano:2012}, suggesting that street networks tend to evolve from a backbone of older streets, and that over time, additional edges connect to this backbone, increasing the network's density. This observation highlights that BC is highly heterogeneous in most street networks, and any reasonable model for these systems should, therefore, capture this specific property. However, we do not require a model to reproduce a specific BC probability distribution, but rather to reflect its heterogeneity. To quantify the heterogeneity in the BC distribution, we calculate the Gini coefficient, $G$, which ranges from $[0,1]$ (see \cite{supplemental}). A value of $G = 0$ corresponds to a uniform BC across all edges, whereas $G = 1$ represents extreme heterogeneity, where a single edge carries high BC while all others have values close to zero. We compute these values for the street network of each urban area in the world, using the street network models from \citep{Boeing:2022}. Our empirical analysis reveals that for most cities worldwide, $G$ is notably large: these networks consistently feature a small number of bottlenecks. Specifically, we find a median $G$ of approximately $0.66$, indicating a pronounced heterogeneity (see Fig.~\ref{fig:fi}(b)). In addition to $G$, we compute the spatial Gini coefficient $G_{\text{spa}}$, which provides a  measure of spatial heterogeneity, by dividing the study area into a regular grid. For each grid cell $c$, we determine the number of network nodes $N_c$ contained within it (the value of the spatial Gini is only weakly affected the grid size, see the SM for details). The spatial Gini coefficient is then calculated from the distribution of these $N_c$ values. $G_{\text{spa}}$ values are also large, with a median around $0.5$ (Fig.~\ref{fig:fi}(c)).

\begin{figure}[bt!]
     \includegraphics[width=0.4\textwidth]{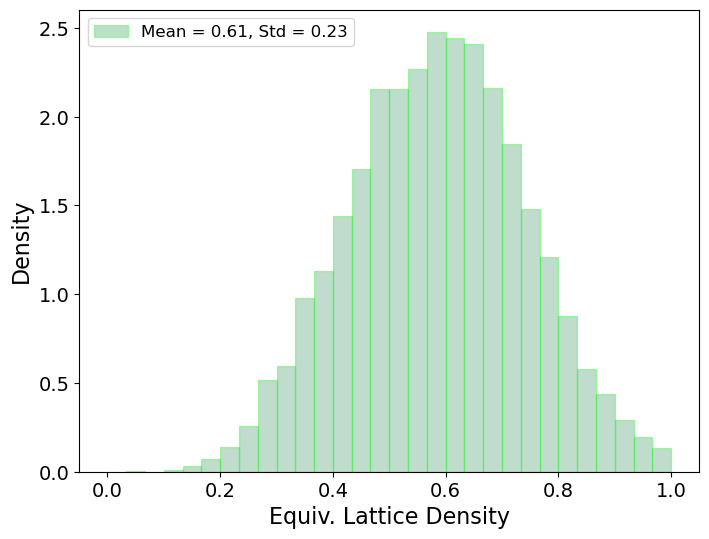}
    \includegraphics[width=0.255\textwidth]{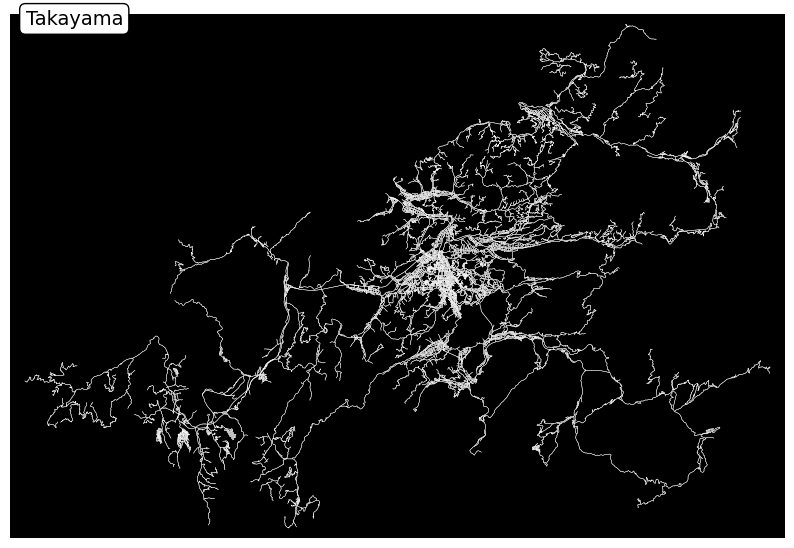}
    \includegraphics[width=0.21\textwidth]{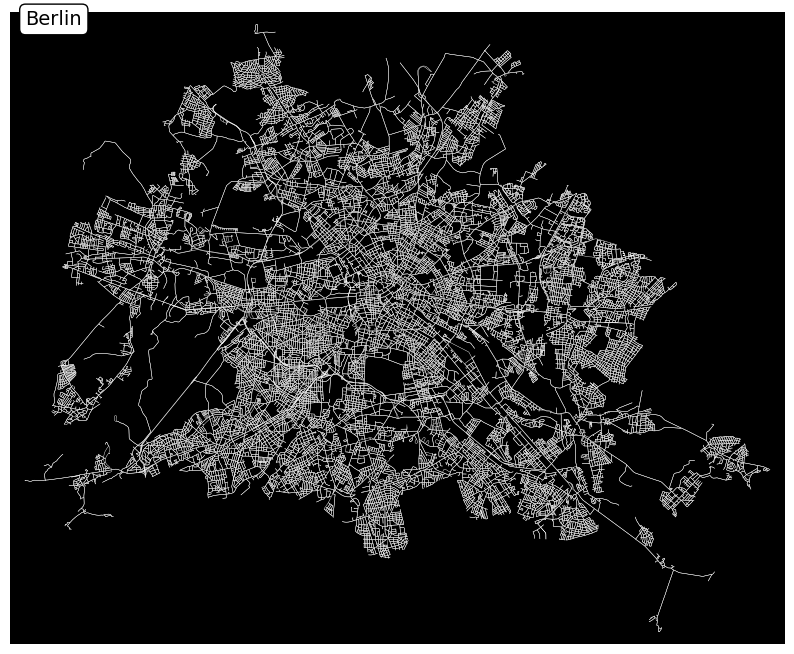}\\
    \includegraphics[width=0.24\textwidth]{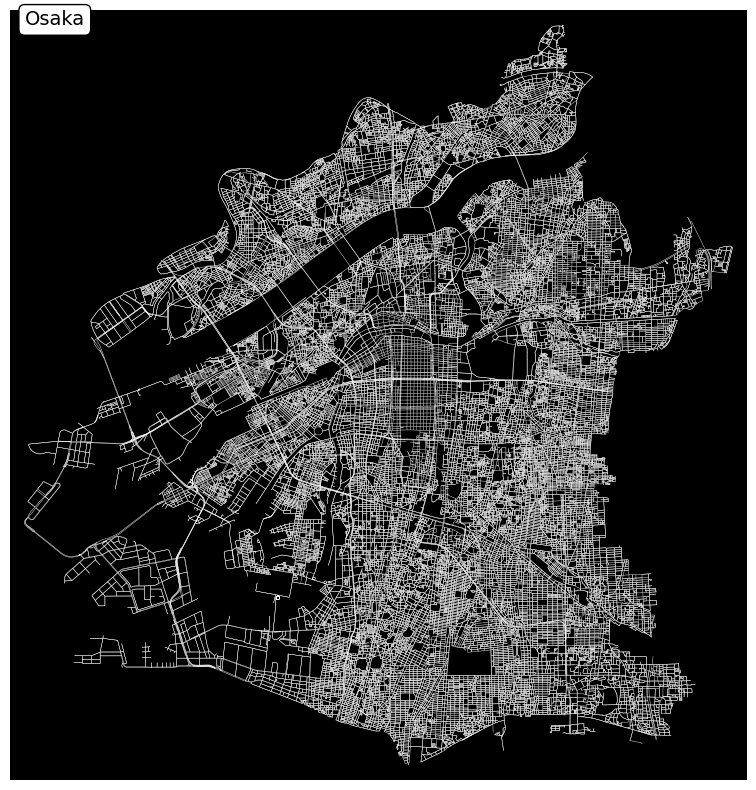}
    \includegraphics[width=0.23\textwidth]{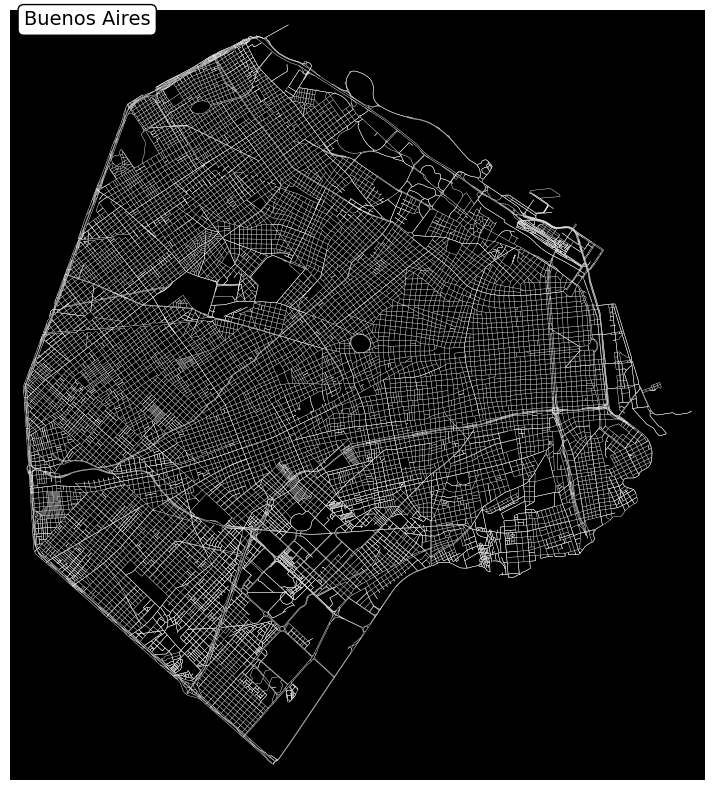}
	\caption{(Top) Equivalent density distribution obtained for about $9,000$ worldwide urban areas. (Bottom) Examples of cities whose street networks exhibit various values of the parameter $p$: Takayama ($p=0.22$), Berlin ($p=0.52$), Osaka ($p=0.69$), Buenos Aires ($p=0.95$). For the method used to compute the value of $p$ see the illustration in \cite{supplemental}.}
	\label{fig:Pp}
\end{figure}

\paragraph*{Equivalent lattice density---}To construct a simple street network model, we map a street network to a grid. Lattice-based networks offer a simplified but powerful representation of real-world systems. If we want to model an equivalent street on a regular grid size we need to estimate the probability that a node of the lattice is present. This {\it equivalent lattice density}, $p$, is estimated in the following way. The number of nodes of a real-world network $N_{\text{emp}}$ represents the
intersections or endpoints in the street network under study (of the city, the urban area or any type of area). We denote by $\ell_1$ the average edge length, and the convex hull area $A$ which is the 
smallest convex polygon that encloses all the nodes in the network (see the SM for more details). If the network were on a regular square lattice, the total length of the square
lattice would then be $\sqrt{A}$ and the grid length would be $\ell_1$. The maximum number of nodes would then be $N_{\max}=(\sqrt{A}/\ell_1)^2=A/\ell_1^2$. The equivalent lattice density is then defined as
\begin{align}
  p=\frac{N_{\text{emp}}}{N_{\max}}=\frac{N_{\text{emp}}\ell_1^2}{A}
\end{align}
This quantity $p$ is dimensionless and combines the density of nodes and the average street length. It is $0\leq p\leq 1$: for $p=1$, the network is equivalent to a regular lattice while for smaller $p$, it is made of smaller clusters that are locally
regular. This parameter provides a standardized way to compare street networks across different cities by accounting for the number of nodes, the average street length, and the area they cover.
Fig.~\ref{fig:Pp} shows the empirical distribution of $p$ around the world, alongside examples of cities with various values of $p$ (other examples can be found in the SM). The average value is $\overline{p}=0.61$ and the standard deviation $\sigma=0.23$.

\begin{figure}[tbh!]
	\includegraphics[width=0.23\textwidth]{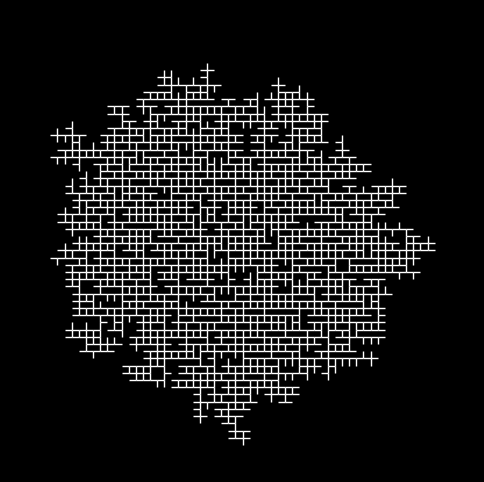}
    \includegraphics[width=0.23\textwidth]{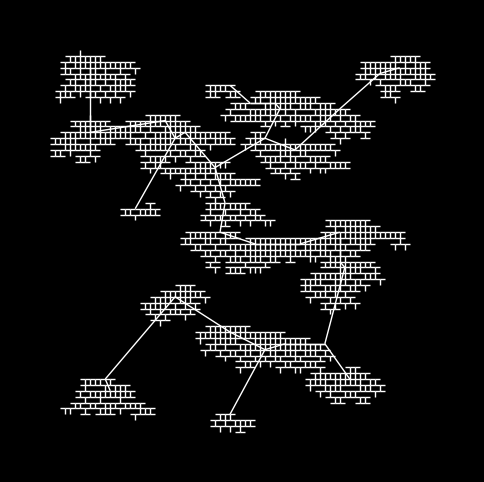}
    \includegraphics[width=0.23\textwidth]{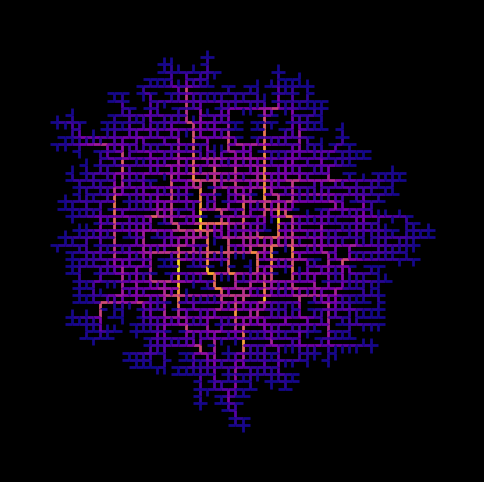}
    \includegraphics[width=0.23\textwidth]{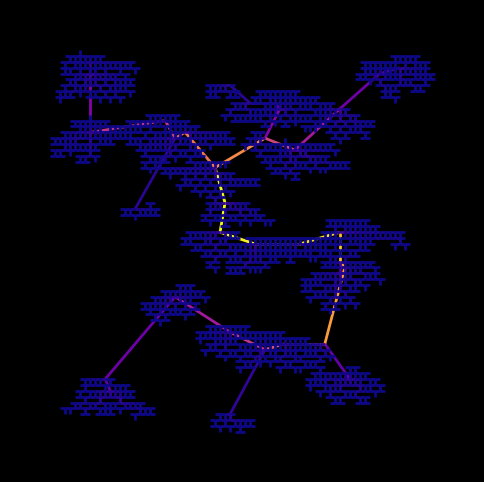}
	\caption{Two different networks with both having an equivalent density $p=0.5$ ($N=4,000$). On the left, we show a Random Eden-like model, and below the corresponding distribution of the Gini coefficient. In this case the Gini coefficient of the BC is $G=0.51$, and the spatial Gini is $G_{spa}=0.26$. On the right, we show the result for our model based on the MST and the corresponding BC map. The BC Gini is here equal to $G=0.72$, and the spatial Gini $G_{\text{spa}}=0.40$.}
	\label{fig:examples}
\end{figure}

\paragraph*{A universal model---}A realistic street network model must match the degree distribution and reproduce key traffic characteristics across values of $p$, notably by capturing the strong heterogeneity observed in BC. This requirement imposes a nontrivial constraint: while it is easy to construct a spatial grid network that satisfies node-level constraints $f_i$, incorporating BC heterogeneity is more difficult. A nearly regular lattice, for example, produces a BC that decays smoothly from the center, yielding low heterogeneity and a small Gini coefficient—far from what is seen in real networks. In order to illustrate this, we used the random Eden model, a simple baseline choice for modeling a road network. This model generates a single connected component through local growth. After the growth phase, the degree proportions $f_i$ can be adjusted to match empirical values (see the SM for details). While other options exist—such as percolating clusters or random geometric graphs—few produce structures with spatial and topological features comparable to real street networks. We show in Fig.~\ref{fig:examples} a network constructed as a random Eden model \cite{eden1961two}. In this case, BC primarily concentrates at the network's center and gradually decreases with distance, resembling the behavior of a regular 2D lattice. Additionally, the heterogeneity remains moderate, with a Gini coefficient of approximately $G \approx 0.5$. As the cluster is compact, the spatial Gini is also low with a value $G_{spa}\approx 0.26$ here. Taking a lattice and removing at random edges (such as in percolation) would lead to a small value of $G$ for the same reasons.

\begin{figure}[tb!]
	\centering
     \includegraphics[width=0.23\textwidth]{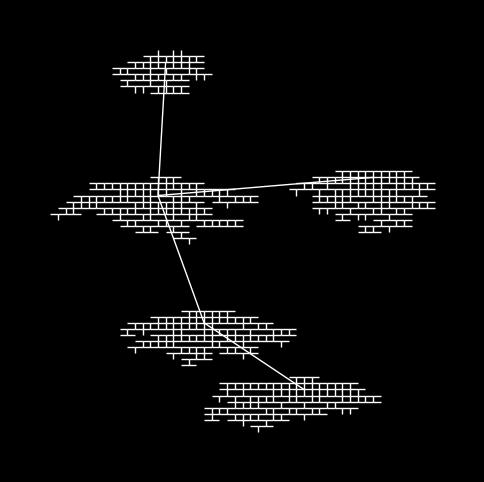}
    \includegraphics[width=0.23\textwidth]{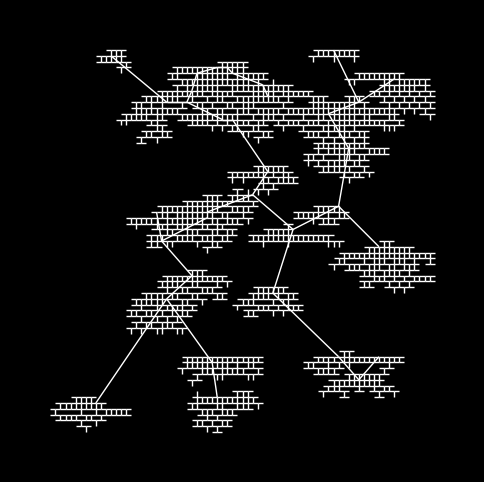}\\
     \includegraphics[width=0.23\textwidth]{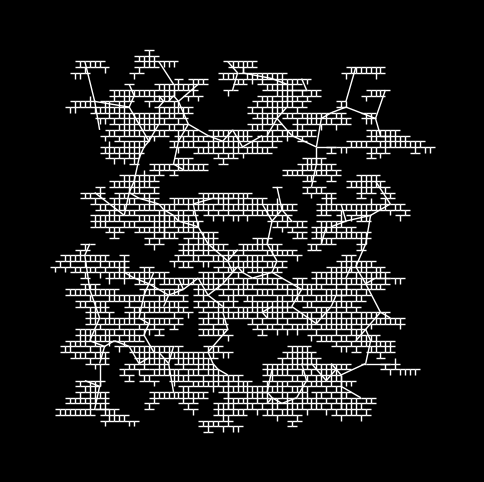}
     \includegraphics[width=0.23\textwidth]{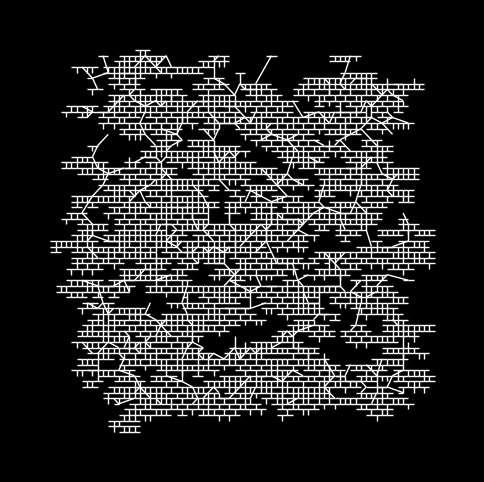}
	\caption{Examples of networks generated by the model for $p=0.23, 0.45, 0.76, 0.92$ and $N=4000$. Target values for the fractions are the empirical ones: $\overline{f_1} = 0.18$, $\overline{f_2} = 0.0$, $\overline{f_3}= 0.62$, and $\overline{f_4} = 0.21$.}
	\label{fig:examples2}
\end{figure}

\begin{figure}[tb!]
	\centering
    \includegraphics[width=0.35\textwidth]{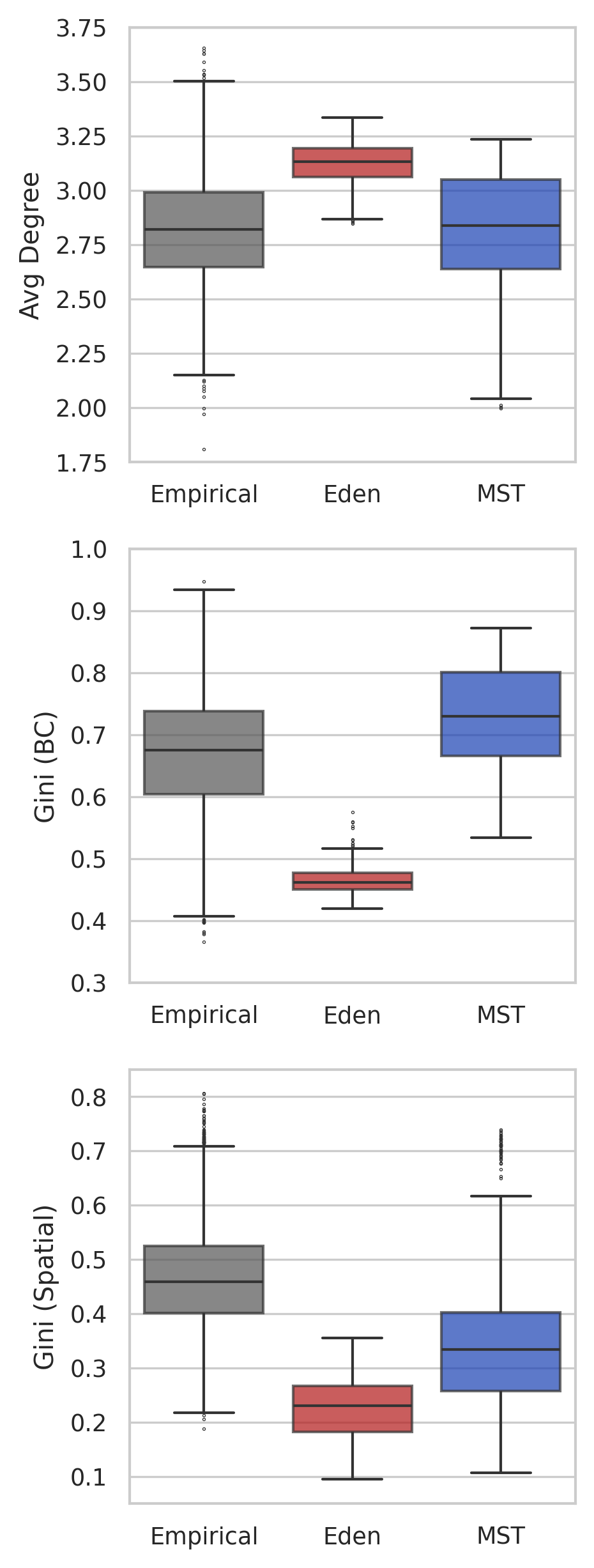}
	\caption{Box plots for the average degree, BC Gini, and spatial Gini values from the empirical data, Eden model, and MST-based model ($N=2,000$ and $50$ configurations for the models). Compared to the Eden model (as a baseline) across a full range of $p$, the MST-based model better reproduces the empirical distributions of these properties' values.}
	\label{fig:box}
\end{figure}

This demonstrates that any compact cluster structure fails to produce sufficiently large BC heterogeneity. To achieve the observed heterogeneity, we follow insights from \cite{Wu:2006, kirkley2018betweenness, barthelemy2022spatial, Barthelemy2024Review}. Our approach begins by constructing a high-BC backbone over a limited subset of nodes, and then incrementally adds edges to match the empirical node degree fractions $f_i$. More precisely, our algorithm (see pseudocode in the SM) starts by randomly selecting a set of grid points and generating a minimum spanning tree (MST) over this set based on Euclidean distances—motivated by results in \cite{Wu:2006}, which highlight the role of MSTs in generating high-BC edges. The network is then iteratively modified through edge additions and deletions to achieve the target degree proportions $f_i$, while ensuring that the total number of nodes converges to a value consistent with the specified grid density $p$, within a given tolerance. To incorporate empirical variability, we define $f_i=\overline{f_i}+\epsilon_i$, where $\overline{f_i}$ is the empirical average and $\epsilon_i$ is a small noise term satisfying $\sum_i\epsilon_i=0$.

Fig.~\ref{fig:examples2} shows examples obtained for different densities $p$ computed with this model. For low densities, the structure exhibits a predominantly modular organization, consisting of small clusters connected by `primary roads' derived from the MST. This configuration sustains a high BC. As density increases, clusters grow and gradually merge. However, the resulting structure does not form a single compact cluster, as seen in models like the Eden growth process for example. Instead, the merged clusters remain interconnected through narrow links, which continue to support a high BC. We observe that these distinct structures can be identified in real-world cities, as illustrated in various cases shown in the SM. 

Once the model converges, we measure various properties--—including the average degree, $G$, and $G_\text{spa}$—--and compare them with empirical data. Fig.~\ref{fig:box} shows that the MST-based model reproduces the empirical properties, providing a good approximation of their distributions, but somewhat under-predicts the spatial Gini coefficient (our simple model uses a regular lattice, whereas real networks exhibit greater spatial flexibility). We also checked that breaking the degeneracy of shortest paths on the lattice doesn't modify our results (see SM). Importantly, however, the MST-based model consistently outperforms the Eden model (as a baseline) at reproducing each of these properties. In \cite{supplemental}, we also present bivariate relationships of each property as a function of $p$, but these are not the focus of the proposed model--—particularly given the dispersion and lack of clear trends in the empirical data themselves.

\paragraph*{Discussion---}This model does not aim to replicate real-world networks in detail but serves as a simplified synthetic proxy, implementing a two-step mechanism: first forming a backbone structure, then incrementally adding edges to achieve a target degree distribution. This approach offers a new method for generating spatial networks and, from a practical perspective, provides a framework for systematically studying how street network structures impact processes like traffic dynamics, CO$_2$ emissions, and beyond. However, the model does not capture certain disordered aspects of real-world networks, such as geographic constraints (e.g., rivers, lakes) or orientational disorder. Despite these limitations, the model successfully explains key features of street networks, including the large heterogeneity in BC (and the creation of high-BC paths) and the significant spatial heterogeneity, as reflected in the large spatial Gini coefficient.

This demonstrates that constructing a network in a naive manner—without integrating the MST process—yields structures that fail to account for the large heterogeneity observed in empirical data. Notably, our model does not require any tunable parameters. With a simple input (such as the size of the lattice and fixed node degree fractions $f_i$ based on average values), it offers a minimal and flexible framework for understanding the structure and properties of street networks. As such, it stands as a promising candidate for a foundational, universal model of street networks.

\bibliography{bibfile_street_network}{}
\bibliographystyle{unsrt}

\section*{Code and data availability}

The python code is freely available and can be downloaded at this address: \url{https://zenodo.org/records/14836480}. All the data is freely available \cite{Boeing:2022}.

\end{document}


\title{Supplemental Material\\ Universal model of urban street networks}

\author{Marc Barthelemy}
\email{marc.barthelemy@ipht.fr}
\affiliation{Universit\'e Paris-Saclay, CNRS, CEA, Institut de Physique Th\'{e}orique, 91191, 
	Gif-sur-Yvette, France}
\affiliation{Centre d'Analyse et de Math\'ematique Sociales (CNRS/EHESS) 54 Avenue de Raspail, 75006 Paris, France}

\author{Geoff Boeing}
\email{}
\affiliation{Department of Urban Planning and Spatial Analysis, Sol Price School of Public Policy, University of Southern California,
  301A Lewis Hall, Los Angeles, CA 90089-0626, USA}

\date{\today}

\maketitle

\section{Data}

The dataset used here comes from [22] and contains originally 8914 urban areas. After cleaning and filtering the data, we are left with 8910 urban areas. It covers urban areas worldwide, using boundaries derived from the Global Human Settlement Layer. Street network data are obtained and modeled from OpenStreetMap with the open-source OSMnx software. In total, the dataset includes over 160 million OpenStreetMap street network nodes and more than 320 million edges across 8,914 urban areas in 178 countries. The geographical distribution is as follows: $55\%$ of these urban areas are in Asia, 
$17\%$ in Africa, $12\%$ in Europe, $11\%$ in Latin America and the Caribbean, $4\%$ in North America, and less than $1\%$ in Oceania.

\section{Gini coefficient}

For a random variable $X$ with realization $x_i$ ($i=1,\dots,n$), the Gini coefficient can be computed as
\begin{align}
    G(X)= \frac{\displaystyle{\sum_{i,j=1}^n \left| x_i - x_j \right|}}{\displaystyle{2 n \sum_{i=1}^n x_i}}  
\end{align}
This quantity is in $[0,1]$. When $G=0$, all the $x_i$ are equal and when $G$ is close to 1, one the $x_i$s is much larger than all the others. 

\section{Impact of the grid size on $G_{\text{spa}}$}

\begin{figure}[h!]
\centering
\includegraphics[width=0.49\linewidth]{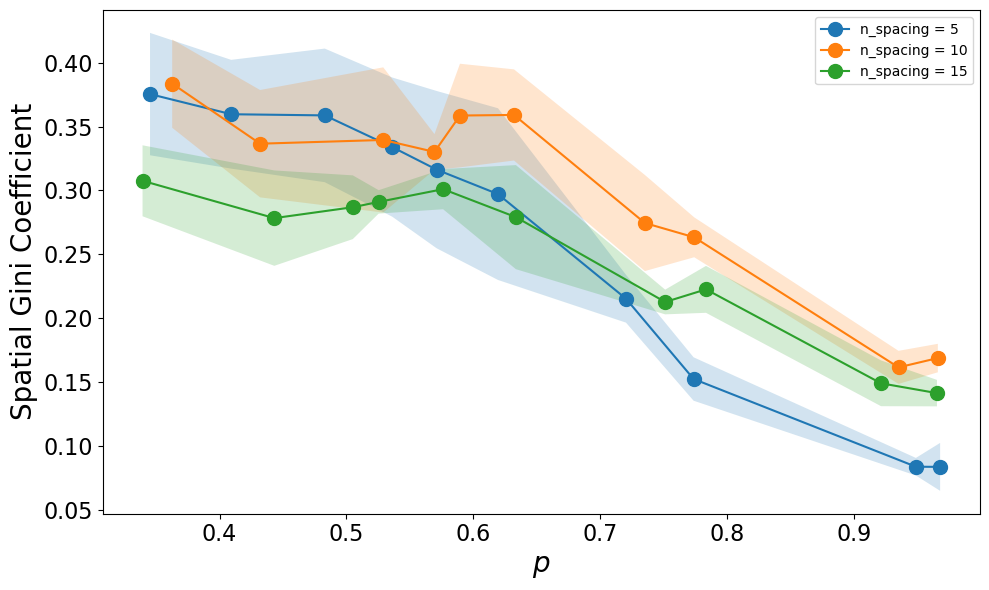}
\caption{Spatial Gini versus $p$ for different values of the spacing size (the area is divided vertically and horizontally by $n_{spacing}$ lines). Results obtained for $N=2,000$ and for $10$ configurations.}
\label{fig:S1}
\end{figure}

The spatial Gini coefficient, $G_{\text{spa}}$, depends a priori on the choice of spatial discretization. To assess the impact of grid resolution, we examine how $G_{\text{spa}}$ varies with the number of divisions used to partition the domain. Specifically, the area was divided into $n_{spacing}$ equal segments along both the horizontal and vertical directions. We then plot $G_{\text{spa}}$ as a function of the effective node fraction $p$ for several values of $n_{spacing}$. The results are presented in Fig.~\ref{fig:S1}.
We observe that the variations in $G_{\text{spa}}$ resulting from changes in $n_{spacing}$ are comparable in magnitude to the fluctuations arising from intrinsic disorder. This suggests that the specific choice of $n_{spacing}$ does not significantly affect the results.

\clearpage
\section{Empirical Measures: Method and Illustration}

\begin{figure}[htp]
	\centering
    \includegraphics[width=0.6\textwidth]{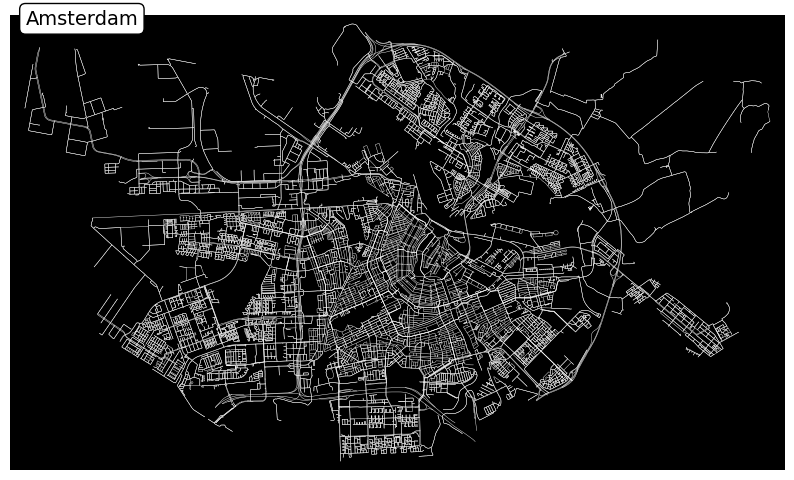}\\
    \centering
    \includegraphics[width=0.6\textwidth]{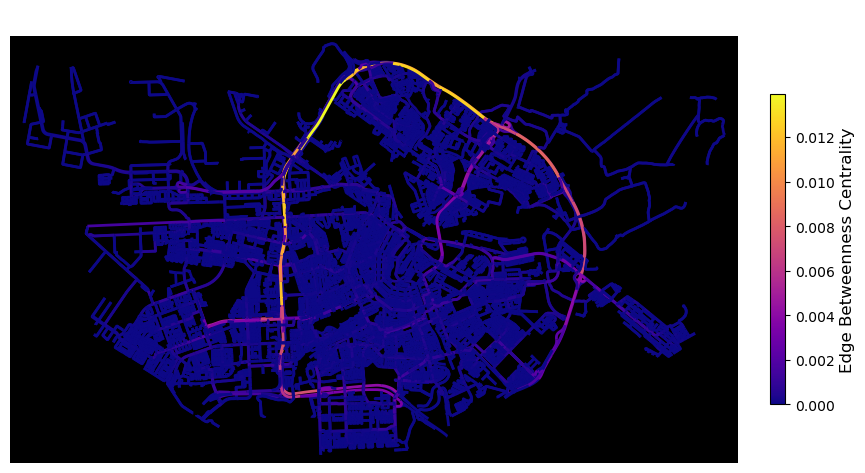}
	\caption{(a) Map of Amsterdam - the data is downloaded from OpenStreetMap. (b) On this network, we then compute the edge betweenness centrality which is shown on the Amsterdam map. From this data, we can then compute the Gini coefficient of the BC.}
	\label{fig:map_amsterdam}
\end{figure}

We illustrate our calculations on the example of the city of Amsterdam (The Netherlands). The data is downloaded using the OSMnX package and the map is shown in Fig.~\ref{fig:map_amsterdam}. We can then compute the edge BC (map also shown in Fig.~\ref{fig:map_amsterdam}), from which we compute the 
Gini coefficient of the BC. 
From this data, we can also make key measures and find for this city, the following values of the fractions:$f_1=0.12$, $f_2=0.02$, $f_3=0.225$, $f_4=0.225$. The average degree is $\overline{k}=3.0$, and the lattice equivalent density is $p=0.56$. The Gini coefficient for the BC is $G=0.84$ - a large value that is reflected in the BC map.

\begin{figure}[htp]
    \includegraphics[width=0.45\textwidth]{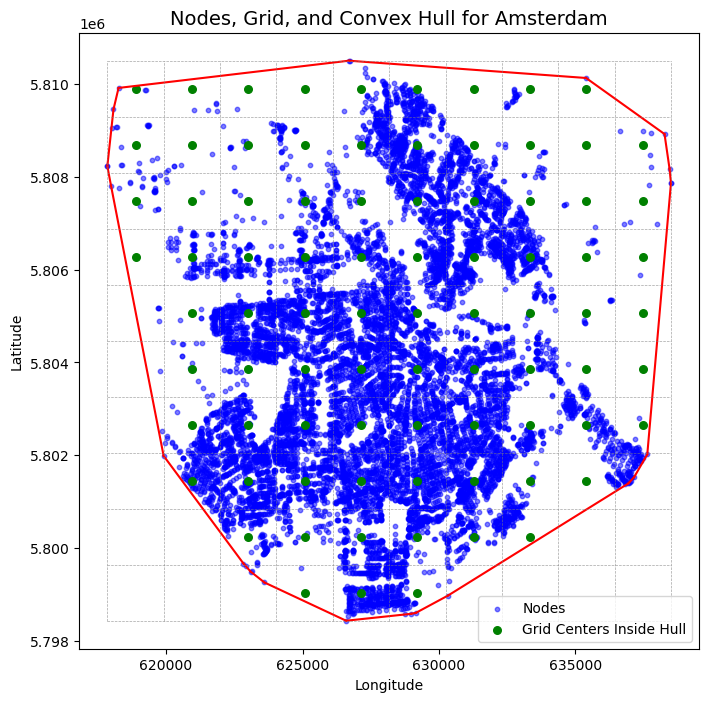}
	\caption{Convex hull for the city of Amsterdam. This allows us to estimate the spatial Gini coefficient.}
	\label{fig:convex}
\end{figure}

In order to compute the spatial Gini coefficient, we determine the convex hull of the city (Fig.~\ref{fig:convex}.
We then divide the space into a grid  and consider grid cells that are inside the convex hull (more precisely those for which the centroid of grid cells are inside the convex hull). For each grid cell $c$, we then compute the number of nodes $N(c)$ belonging to it, and from this, compute the Gini coefficient of these $N(c)$.  The spatial Gini coefficient is here $G_{spa}=0.50$, a moderate value indicating that disparities are not very large.

\newpage
\clearpage
\section{Examples of cities with various values of $p$}

\begin{figure}[htp]
    \includegraphics[width=0.25\textwidth]{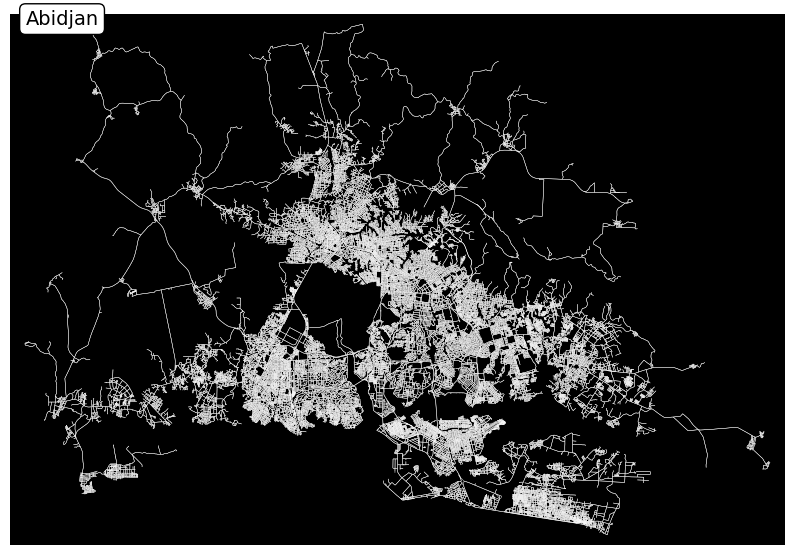}
    \includegraphics[width=0.20\textwidth]{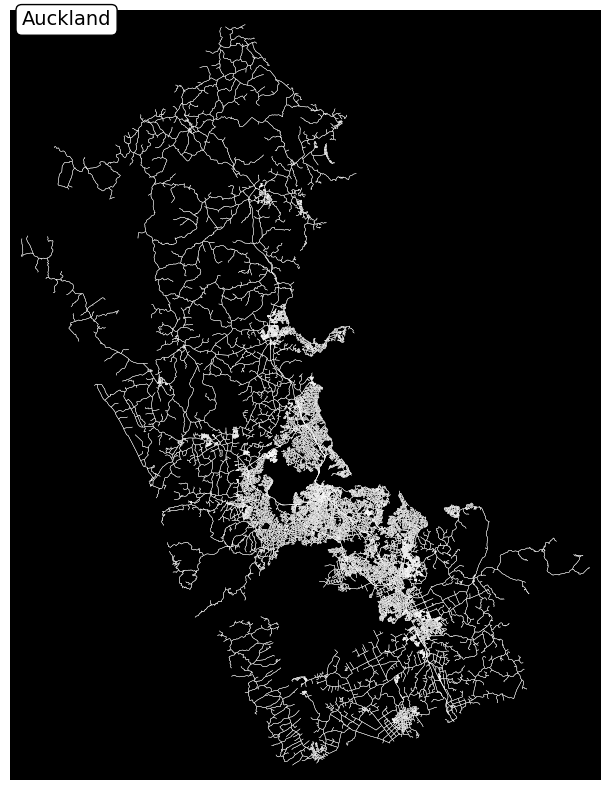}
    \includegraphics[width=0.25\textwidth]{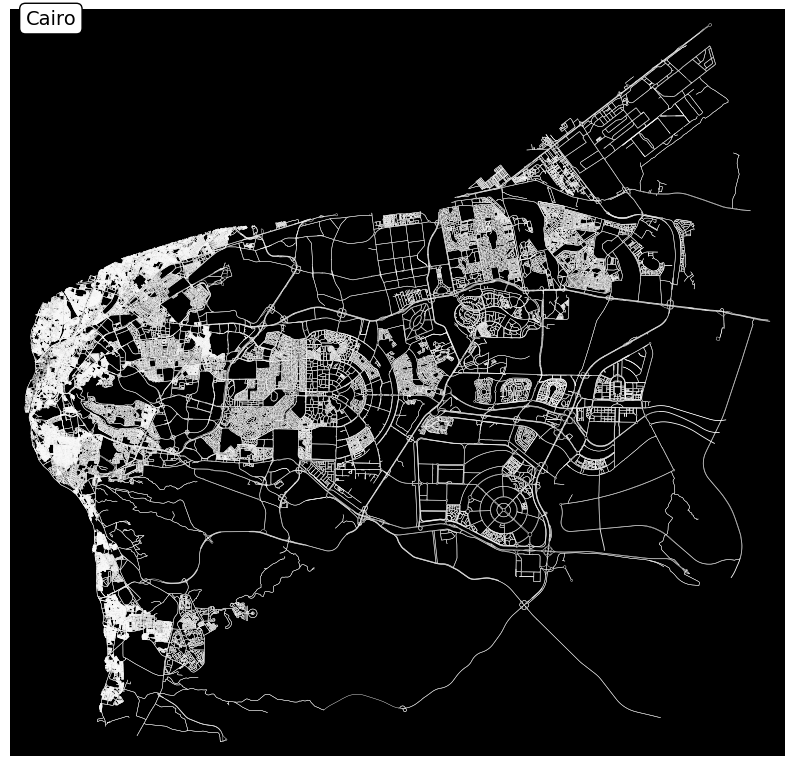}
    \includegraphics[width=0.25\textwidth]{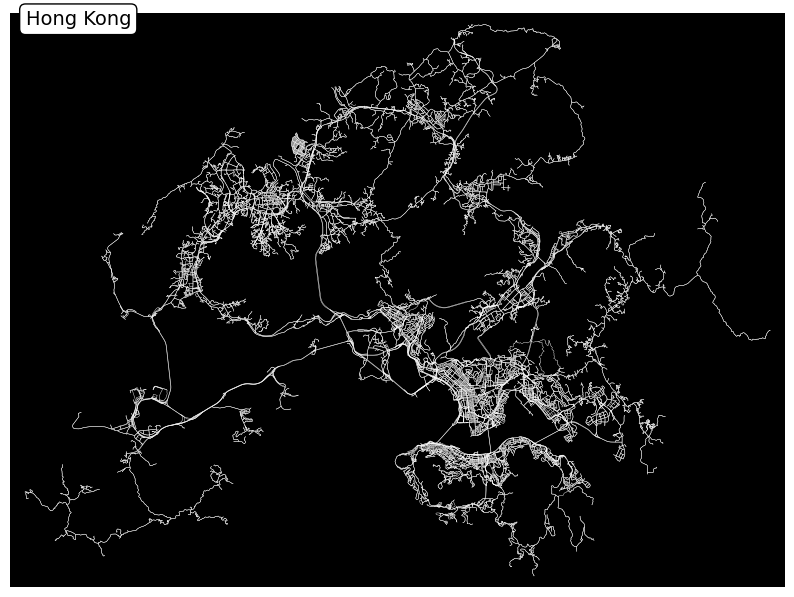}
	\caption{Low density values cities: Abidjan ($p=0.29$), Auckland ($p=0.18$), Cairo ($p=0.26$), Hong Kong (Island) ($p=0.29$).}
	\label{fig:lowvalues}
\end{figure}

\begin{figure}[htp]
    \includegraphics[width=0.25\textwidth]{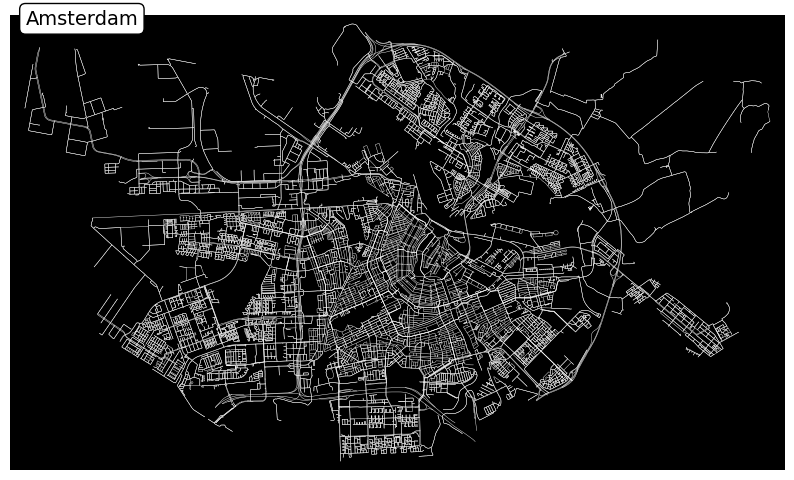}
    \includegraphics[width=0.25\textwidth]{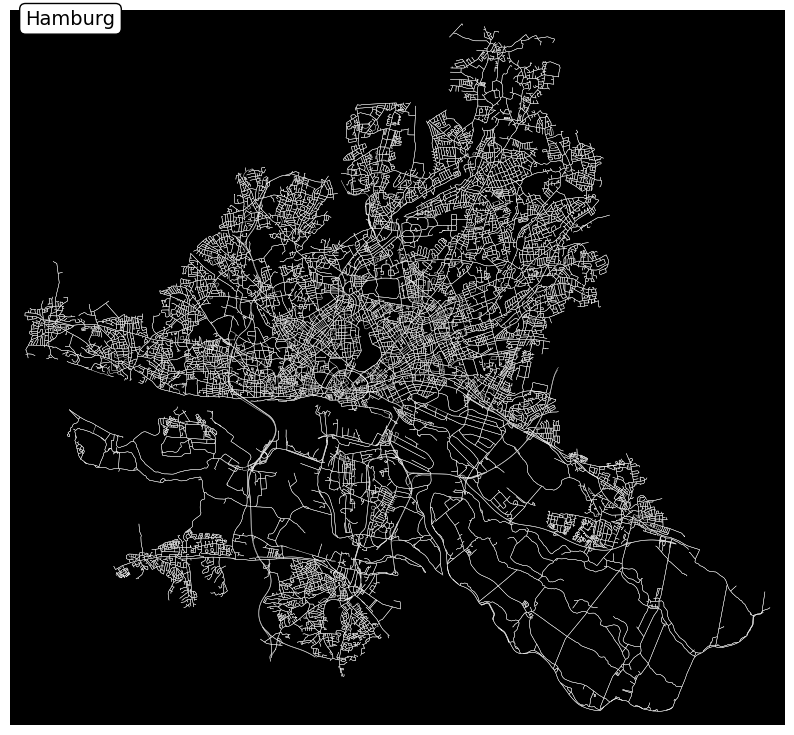}
    \includegraphics[width=0.25\textwidth]{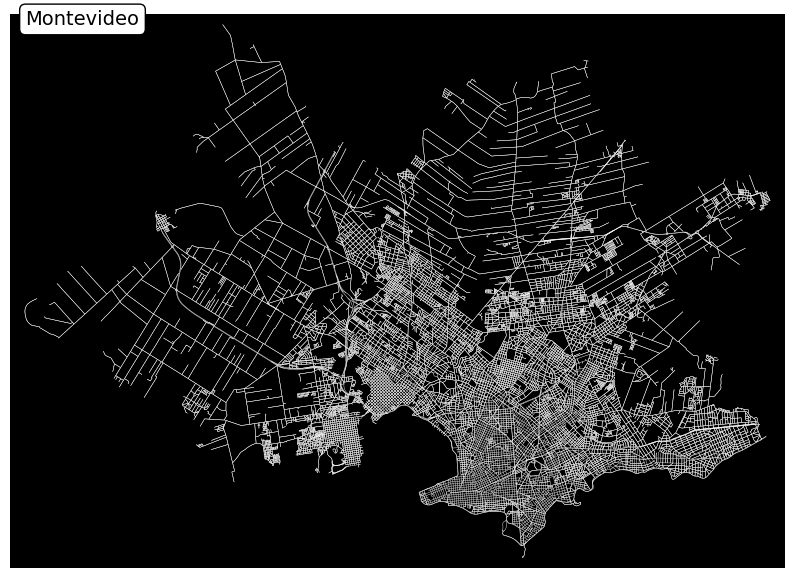}
    \includegraphics[width=0.20\textwidth]{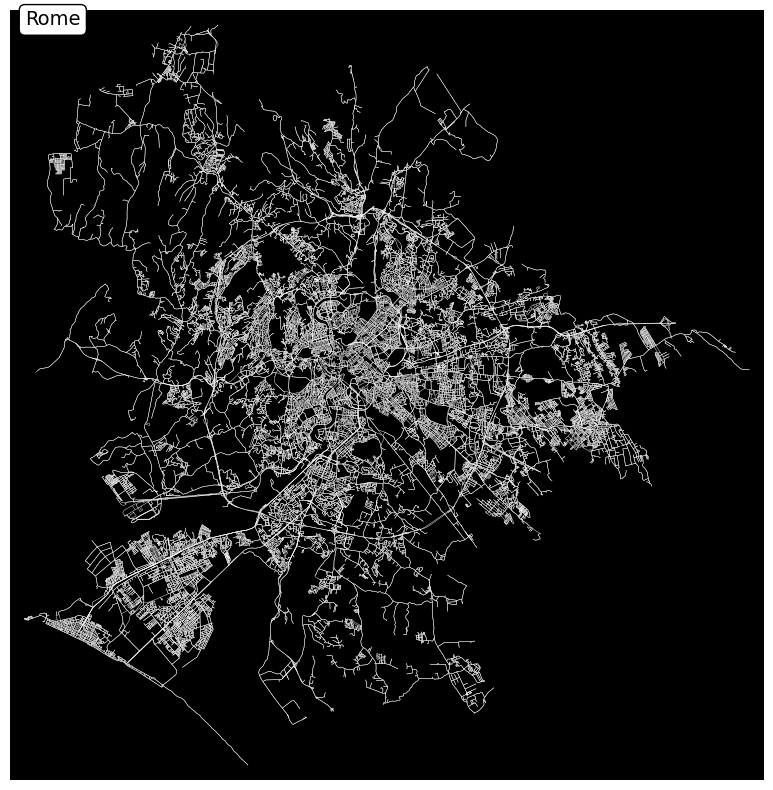}
	\caption{Cities with $p~\sim 0.5$: Amsterdam ($p=0.55$), Hamburg ($p=0.49$), Montevideo ($p=0.44$), Rome ($p=0.41$).}
	\label{fig:midvalues}
\end{figure}

\begin{figure}[htp]
    \includegraphics[width=0.25\textwidth]{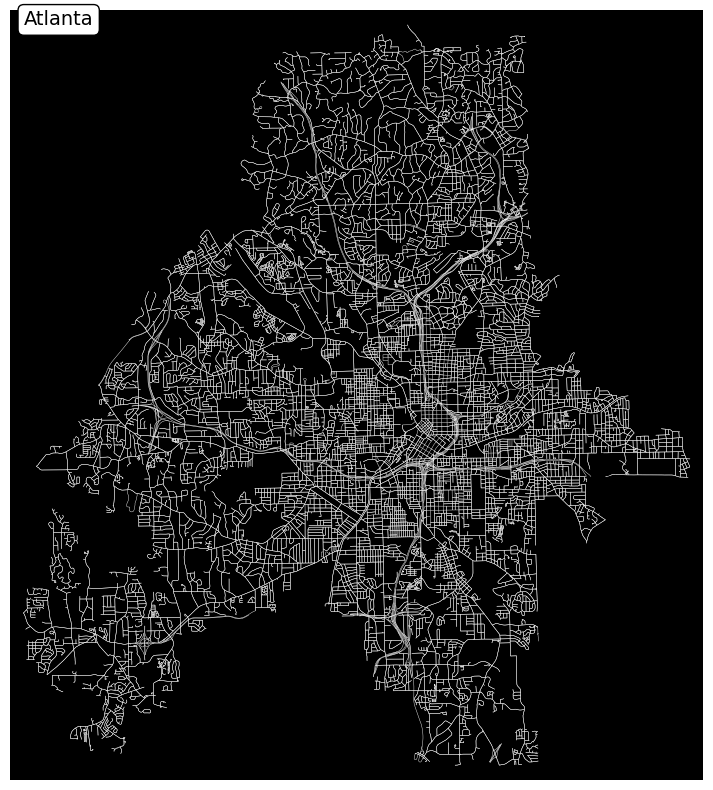}
    \includegraphics[width=0.20\textwidth]{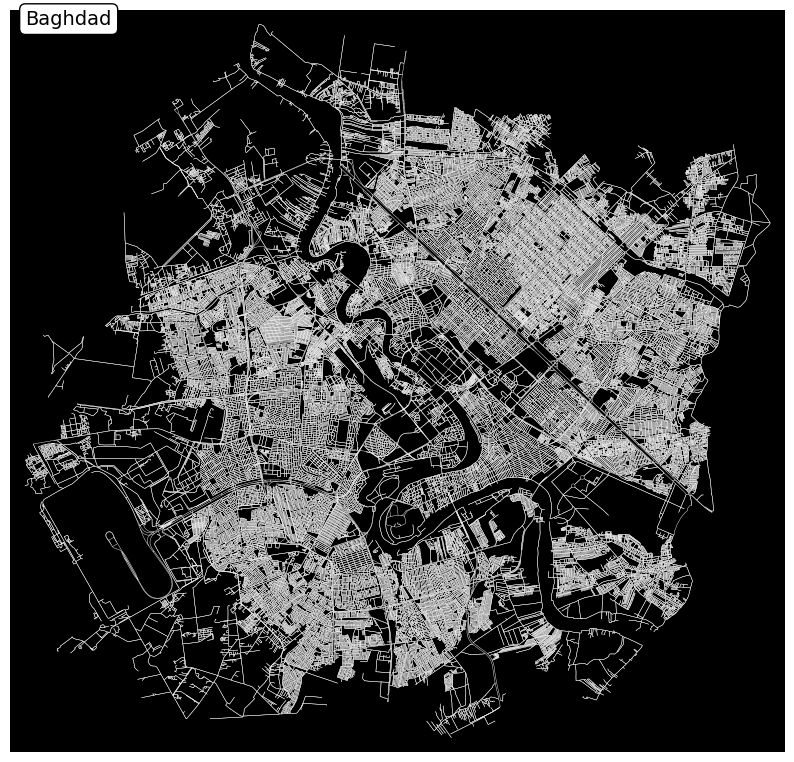}
    \includegraphics[width=0.25\textwidth]{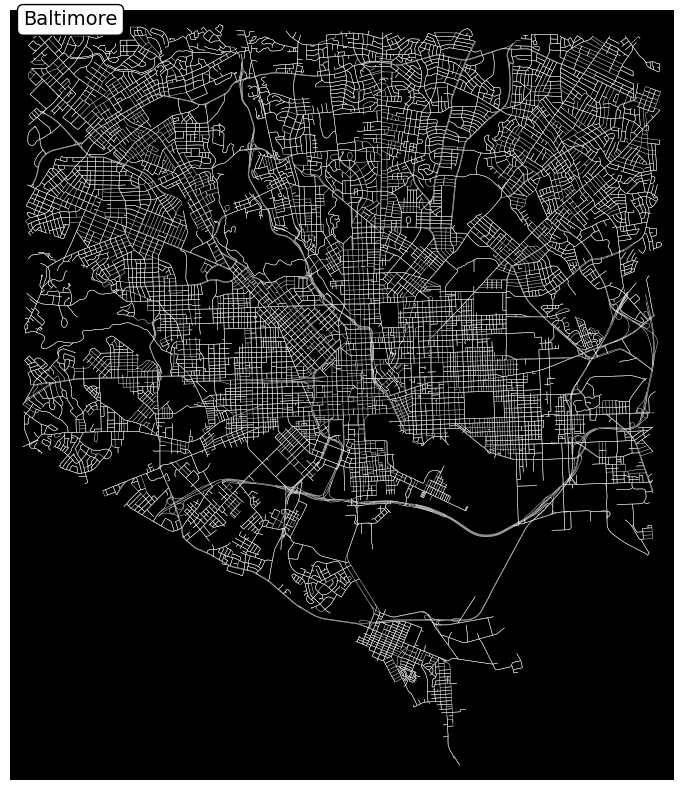}
    \includegraphics[width=0.25\textwidth]{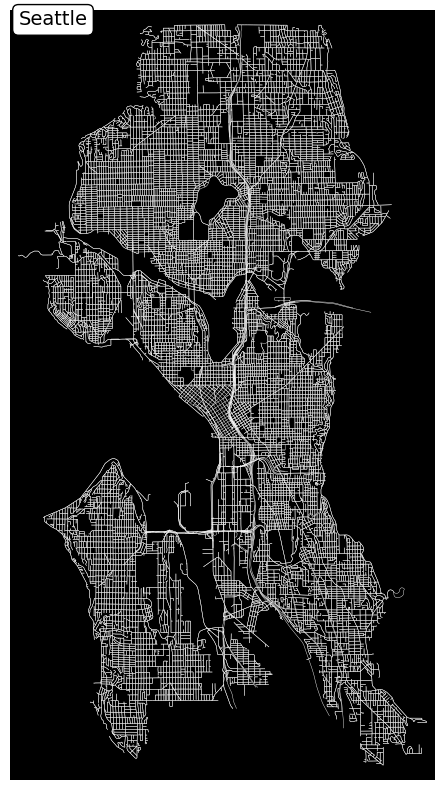}
	\caption{Cities with $p\sim 0.75$: Atlanta ($p=0.67$), Baghdad ($p=0.73$), Baltimore ($p=0.78$), Seattle ($p=0.74$).}
	\label{fig:midplusvalues}
\end{figure}

\begin{figure}[htp]
    \includegraphics[width=0.25\textwidth]{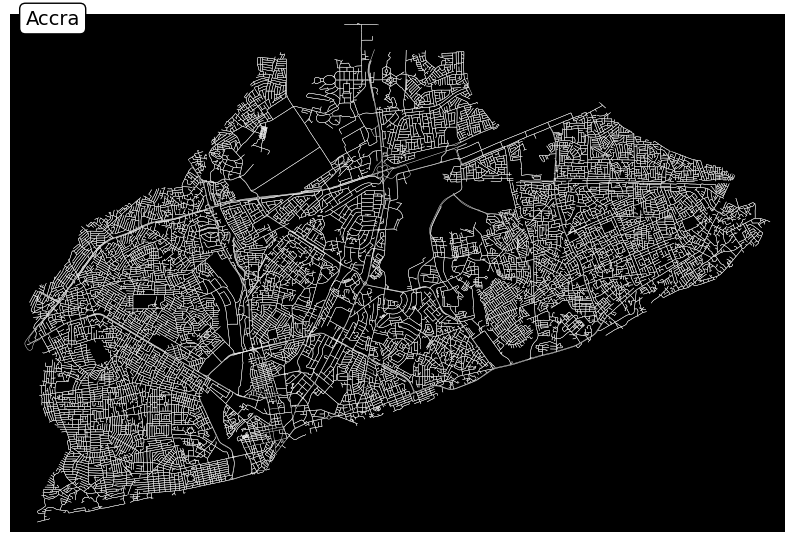}
    \includegraphics[width=0.20\textwidth]{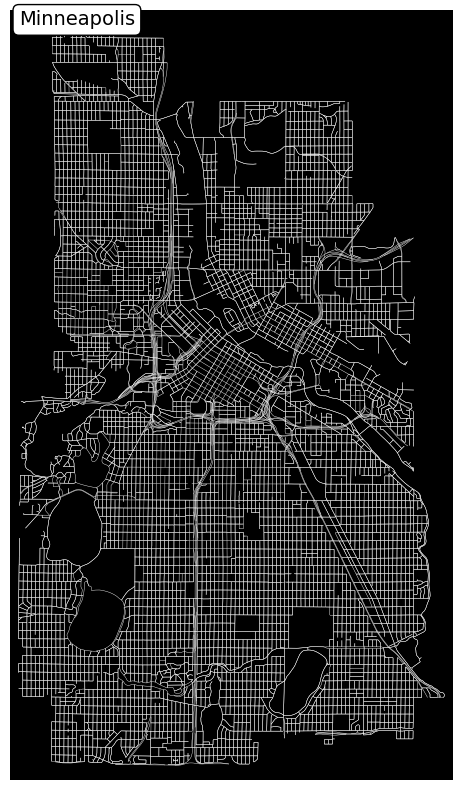}
    \includegraphics[width=0.25\textwidth]{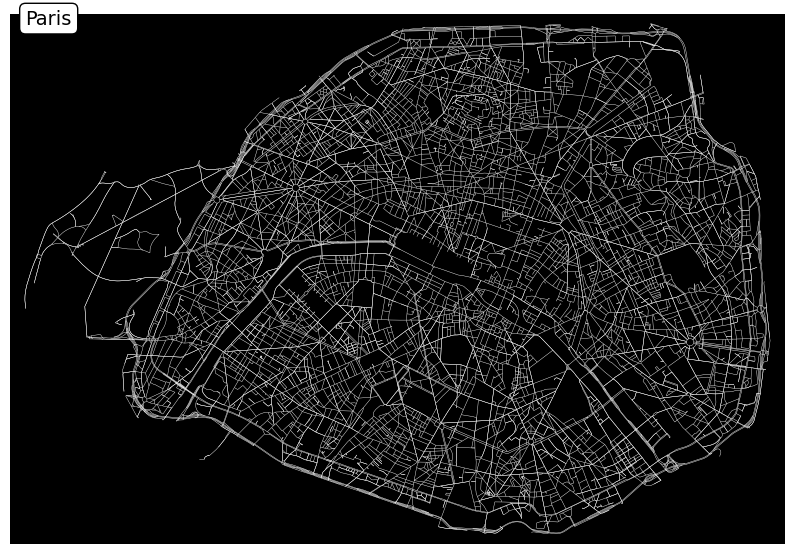}
    \includegraphics[width=0.25\textwidth]{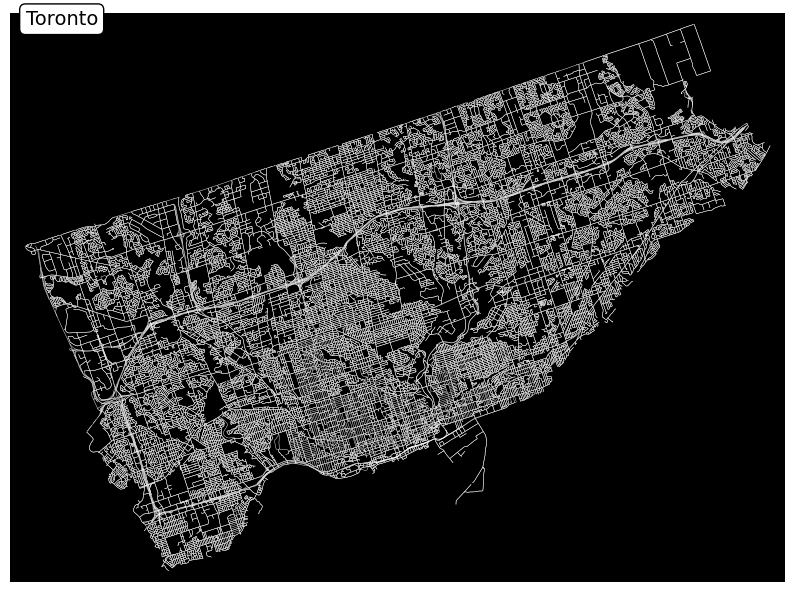}
	\caption{Cities with $p$ close to 1: Accra ($p=0.94$), Minneapolis ($p=0.99$), Paris ($p=0.91$), Toronto ($p=0.89$).}
	\label{fig:highvalues}
\end{figure}

We show here 4 different `families' of cities with low values of $p~0.2-0.3$ (Fig.~\ref{fig:lowvalues}), 
intermediate values of order $p\sim 0.5$ (Fig.~\ref{fig:midvalues}), 
larger values of order $p\sim 0.75$ (Fig.~\ref{fig:midplusvalues}), 
and large values close to 1 (Fig.~\ref{fig:highvalues}).

\clearpage
\newpage
\section{Eden-like model: Code description}

\section*{Subroutine: Create Eden-Like Connected Component from Grid Center}

\textbf{Input:}
\begin{itemize}
  \item $N_{\text{init}}$: Target number of nodes
  \item $grid\_points$: List of all available positions on the grid
  \item $grid\_side$: Number of grid units along one side (for bounds)
\end{itemize}

\vspace{1ex}
\textbf{Steps:}
\begin{tabbing}
1. \= Define the central point: $center = (\lfloor grid\_side/2 \rfloor, \lfloor grid\_side/2 \rfloor)$ \\
2. \= Initialize graph $G$ with the center node \\
3. \= Initialize $added\_nodes$ with the center node \\
4. \= Initialize $potential\_neighbors$ with the 4 nearest neighbors of the center \\
5. \= \textbf{While} $|added\_nodes| < N_{\text{init}}$ and $potential\_neighbors$ is not empty: \\
\> a. \= Randomly select $new\_node$ from $potential\_neighbors$ \\
\> b. \= Remove $new\_node$ from $potential\_neighbors$ \\
\> c. \= Identify neighbors of $new\_node$ that are already in $added\_nodes$ \\
\> d. \= If such neighbors exist: \\
\>\> i. \= Randomly select one $existing\_node$ among them \\
\>\> ii. \= Add $new\_node$ to $G$ and connect it to $existing\_node$ \\
\>\> iii. \= Add $new\_node$ to $added\_nodes$ \\
\>\> iv. \= For each of its 4-grid neighbors, if valid and not already in $added\_nodes$, add to $potential\_neighbors$
\end{tabbing}

\vspace{1ex}
\textbf{Output:} A graph $G$ with positions $pos$ representing an Eden-like connected component

\clearpage
\newpage
\section{Model: Code description}

In order to reproduce empirical fluctuations, the code described here is used for generating networks with fraction given by $f_i = \overline{f_i} + \varepsilon_i$ where $\overline{f_i}$ is the empirical observed average and $\varepsilon_i$ is small noise such that $\sum_i\varepsilon_i=0$.

\section*{Subroutine 1: Create Connected Component from Random Points}

\textbf{Input:} $N_{\text{mst}}$ (number of nodes), $grid\_points$ (available positions)

\vspace{1ex}
\textbf{Steps:}
\begin{tabbing}
1. \= Randomly select $N_{\text{mst}}$ points from $grid\_points$ and assign to $random\_points$ \\
2. \= Create graph $G$ with $random\_points$ as nodes \\
3. \= Compute pairwise Euclidean distances between all node pairs \\
4. \= Add weighted edges to $G$ using these distances \\
5. \= Compute the minimum spanning tree $mst$ of $G$
\end{tabbing}

\vspace{1ex}
\textbf{Output:} $mst$ (minimum spanning tree), $random\_points$ (selected points)

\vspace{2em}

\section*{Subroutine 2: Compute Degree Statistics}

\textbf{Input:} $G$ (graph)

\vspace{1ex}
\textbf{Steps:}
\begin{tabbing}
1. \= Compute degree of each node in $G$ \\
2. \= Count number of nodes with degrees 1, 2, 3, and 4 \\
3. \= Compute degree fractions $f_1$, $f_2$, $f_3$, $f_4$ by normalizing counts \\
4. \= Compute average degree as a weighted sum of degrees
\end{tabbing}

\vspace{1ex}
\textbf{Output:} Degree fractions $f_1$, $f_2$, $f_3$, $f_4$ and average degree

\vspace{2em}

\section*{Subroutine 3: Compute Gini Coefficient}

\textbf{Input:} $x$ (array of values)

\vspace{1ex}
\textbf{Steps:}
\begin{tabbing}
1. \= Sort $x$ in ascending order \\
2. \= Compute cumulative sums of sorted values \\
3. \= Compute the Gini coefficient using the standard formula
\end{tabbing}

\vspace{1ex}
\textbf{Output:} $gini$ (Gini coefficient)

\vspace{2em}

\section*{Subroutine 4: Compute Spatial Gini Coefficient}

\textbf{Input:}
\begin{itemize}
  \item $positions$: list of $(x, y)$ coordinates of nodes
  \item $grid\_side$: side length of the spatial domain
  \item $n\_spacing$: number of cells per side (grid resolution)
\end{itemize}

\vspace{1ex}
\textbf{Steps:}
\begin{tabbing}
1. \= Compute $cell\_size = grid\_side / n\_spacing$ \\
2. \= Initialize a 2D array $counts$ of zeros with shape $(n\_spacing, n\_spacing)$ \\
3. \= For each node position $(x, y)$ in $positions$: \\
\> a. \= Determine the cell index: \\
\>\> i. \= $cell\_x = \min(\lfloor x / cell\_size \rfloor, n\_spacing - 1)$ \\
\>\> ii. \= $cell\_y = \min(\lfloor y / cell\_size \rfloor, n\_spacing - 1)$ \\
\> b. \= Increment the count in $counts[cell\_x, cell\_y]$ \\
4. \= Flatten the $counts$ array to a 1D vector $counts\_flat$ \\
5. \= Extract non-zero values: $counts\_nonzero = \{c \in counts\_flat \mid c > 0\}$ \\
6. \= If $counts\_nonzero$ is empty, return 0 \\
7. \= Otherwise, return $\text{Gini}(counts\_nonzero)$
\end{tabbing}

\vspace{1ex}
\textbf{Output:} Spatial Gini coefficient

\vspace{2em}

\section*{Subroutine 5: Optimize Network}

\textbf{Input:} $p$, $N_0$, $N_{\text{mst}}$, $f_{t1}$, $f_{t2}$, $f_{t3}$, $f_{t4}$ (target degree fractions), \\
$tolerance$, $size\_tolerance$, $n\_spacing$, $max\_iterations$

\vspace{1ex}
\textbf{Steps:}
\begin{tabbing}
1. \= Create square grid with $N_0$ nodes \\
2. \= Compute target node count $N = p \cdot N_0$ \\
3. \= Initialize MST with $N_{\text{mst}}$ nodes \\
4. \= For each iteration up to $max\_iterations$: \\
\> a. \= Compute current degree fractions $f_1$, $f_2$, $f_3$, $f_4$ and node count \\
\> b. \= If all $|f_i - f_{ti}| < tolerance$ and node count is within $size\_tolerance$, break \\
\> c. \= Otherwise: \\
\>\> i. \= If $f_4 < f_{t4}$, select a degree-3 node and add an edge \\
\>\> ii. \= Else if $f_3 < f_{t3}$, select a degree-2 node and add an edge \\
\>\> iii. \= Else, select a degree-1 node and add two edges
\end{tabbing}

\vspace{1ex}
\textbf{Output:} Optimized graph $G$

\vspace{2em}

\section*{Subroutine 6: Compute Final Network Metrics}

\textbf{Input:} Same parameters as in Subroutine 5

\vspace{1ex}
\textbf{Steps:}
\begin{tabbing}
1. \= Optimize network using Subroutine 5 \\
2. \= Compute degree statistics using Subroutine 2 \\
3. \= Compute betweenness centrality Gini coefficient using Subroutine 3 \\
4. \= Compute spatial Gini coefficient using Subroutine 4
\end{tabbing}

\vspace{1ex}
\textbf{Output:} Degree fractions, BC Gini coefficient, spatial Gini coefficient

\clearpage
\section{Gini coefficient for the Eden-like model}

\begin{figure}[htp]
    \includegraphics[width=0.7\textwidth]{./
    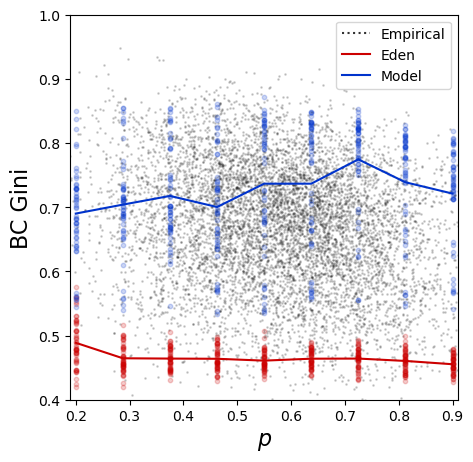}
	\caption{BC Gini coefficient $G$ versus $p$ for the Eden-like model ($N=2000$, $50$ configurations), compared to the MST-based model and the empirical results for $N\sim 9000$ urban areas worldwide.}
	\label{fig:eden}
\end{figure}

We show on Fig.~\ref{fig:eden} the Gini coefficient $G$ versus $p$ computed for the Eden-like model ($N=2000$, $50$ configurations), and compared to the empirical results for $N\sim 9000$ urban areas worldwide.
We observe that the results for the Eden-like model yield a Gini coefficient of approximately $G\sim 0.5$, which is significantly lower than the empirical values (ranging roughly from 0.75 to 0.6). In contrast, the improvement achieved by our MST-based model is evident, as it successfully reproduces the large BC Gini values observed in the empirical data.

\clearpage
\section{Analysis of the Gini coefficients vs. $p$}

We observe a slight decreasing trend in the BC Gini as a function of $p$, which can be understood as follows: as the network becomes denser, the number of alternate routes increases, reducing the betweenness centrality of the original MST segments and thereby globally decreasing the BC heterogeneity. For the same reason, we also observe a decreasing trend in the spatial Gini coefficient: when networks become denser, with more nodes, there is a natural trend towards a homogeneous distribution, thereby decreasing the spatial gini coefficient. 
\begin{figure}[h!]
\centering
\includegraphics[width=0.9\linewidth]
{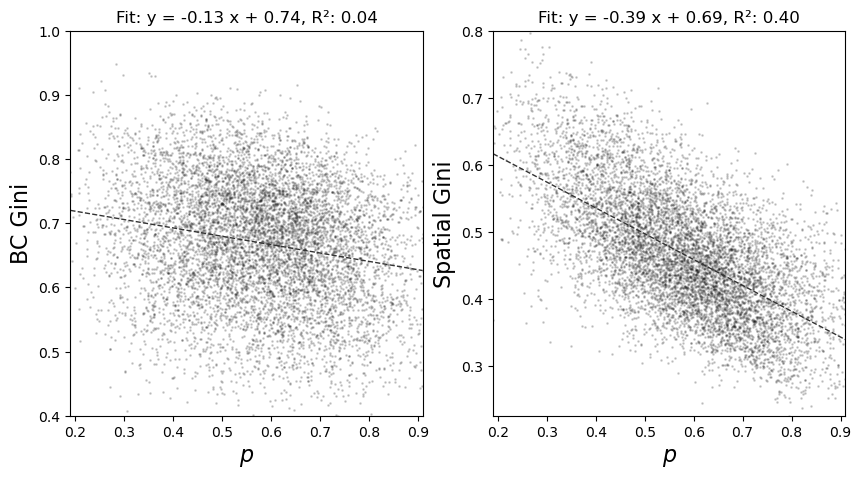}
\caption{Plot of the BC Gini coefficient for the Eden like model versus p and compared to both the model and the empirical data. The results were obtained for $N=2,000$ and for $50$ configurations. }
\label{fig:trends}
\end{figure}

Despite these two similar trends (for the BC and spatial Gini coefficients), they are however not correlated to each other. In order to test for the correlation between these two Gini coefficients, we compute the Pearson correlation coefficient for each value of $p$: $r(p)=\text{PearsonCorrelation}({G_{BC}(p)},{G_{spa}(p)})$. The results are shown in the following Table~\ref{tab:correlation_gini}.
\begin{table}[h!]
\caption{Pearson correlation between spatial Gini and BC Gini per $p$ (obtained for $1,000$ configurations and $N=2,000$).}
\label{tab:correlation_gini}
\centering
\begin{tabular}{rrr}
\toprule
$p$ & Pearson $r(p)$ & p-value \\
\midrule
0.20 & -0.053 & 0.092 \\
0.375 & -0.028 & 0.374 \\
0.55 & -0.039 & 0.219 \\
0.725 & -0.020 & 0.535 \\
0.90 & -0.019 & 0.557 \\
\bottomrule
\end{tabular}
\end{table}
We observe that the Pearson correlation coefficients are small, and the corresponding p-values are large. These results indicate that the correlations are not statistically significant, suggesting that the two Gini coefficients are not linearly correlated and instead capture independent structural properties of the network. This is expected, as they represent fundamentally different aspects of the network: the BC Gini reflects the heterogeneity of flow across the network, while the spatial Gini quantifies the spatial distribution of the nodes.

\begin{figure}[h!]
\centering
\includegraphics[width=0.7\linewidth]{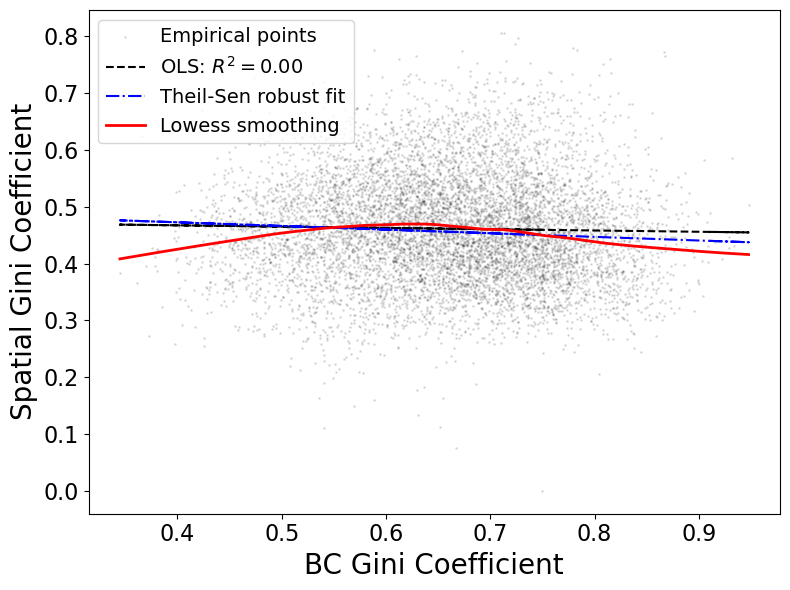}
\caption{{\bf Correlations BC-spatial Gini.} Spatial Gini versus BC Gini (for all values of $p$) together with various regression methods.}
\label{fig:corre}
\end{figure}

We also show (Fig.~\ref{fig:corre}) the Spatial versus the BC Gini for all values of $p$. Various regression methods can be applied and show not significative correlations. 

\clearpage
\section{Bivariate relations and comparison between models and empirical data}

\begin{figure*}[htp]
\centering
     \includegraphics[width=0.9\textwidth]{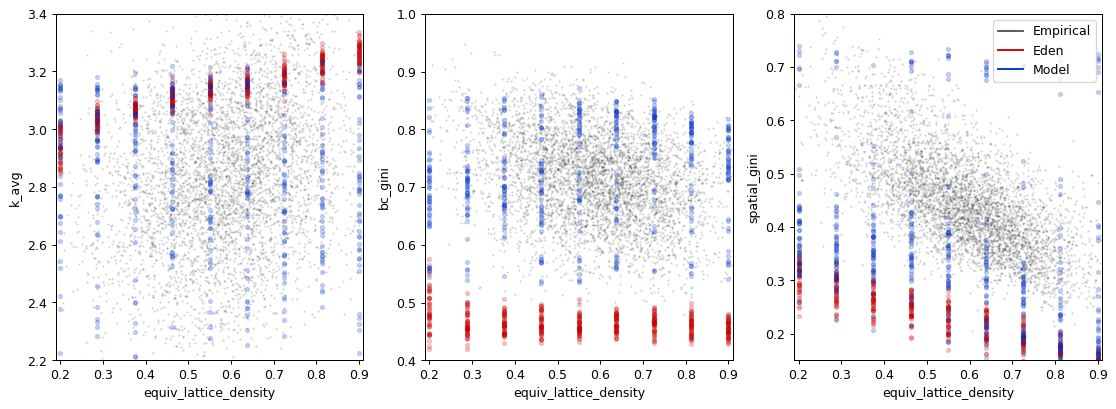}
\caption{Comparison of model's predictions and empirical results measured on $\sim 9,000$ worldwide cities. (Left) Average degree versus $p$. (Middle) Gini coefficient for the BC, (Right) spatial Gini coefficient. In these figures, we also display the results for the Eden model, showing that a simple model cannot explain the large heterogeneities observed in the data.  The theoretical results are obtained for $N=2,000$ nodes, and for $100$ configurations.}
	\label{fig:all}
\end{figure*}

We present in this section the comparison of all the indicators versus $p$. The results are shown in Fig.~\ref{fig:all}. 


We observe that our model successfully generates networks with high betweenness centrality (BC) and spatial Gini coefficients. Moreover, by design, the model accurately reproduces the average degree of the network. For the BC Gini coefficient (Fig.~\ref{fig:all}(Middle)), the predicted values are generally consistent with empirical data, although there is a slight tendency toward overestimation, and the model fails to reproduce the modest decreasing trend observed in the data. This decrease can be interpreted as a consequence of increasing network density: as more connections are added, the number of alternative paths grows, reducing the betweenness centrality of previously important segments. This effect is not captured by the model, in which the original MST segments retain high BC values even at large $p$. While our results demonstrate that the MST-based model can reproduce high Gini coefficients, it does not account for this subtle decline. Nonetheless, the primary goal of the model was to generate networks with strong BC heterogeneity using a simple mechanism. In contrast, we find that a simpler, "naive" model such as the Eden model fails to reproduce the large Gini values observed empirically. \\

The variation of the spatial Gini coefficient with $p$ (Fig.~\ref{fig:all}(Right)) follows, on average, the empirically observed decreasing trend. This behavior can be intuitively understood: in denser networks, it becomes increasingly difficult to form structures with pronounced heterogeneities, such as small clusters connected by only one or a few roads (note that the spatial Gini coefficient is sensitive to the choice of grid spacing; we tested different values and observed similar results, see part III of this Supp. Mat. for more details). Although the decreasing trend with $p$ is consistently reproduced, the model does not reach the high spatial Gini values observed in empirical data. Nevertheless, it still represents a clear improvement over the Eden-like model. This discrepancy is likely due to the fact that our model is built on a regular lattice, whereas real-world networks are less constrained and can exhibit greater spatial heterogeneity. One possible way to address this limitation would be to adjust the geometric coordinates in the model; however, doing so would significantly increase its complexity, potentially undermining its simplicity and generality.

\clearpage
\section{Shortest paths degeneracy}

On a square lattice, shortest paths are degenerate—this is accounted for in the definition of betweenness centrality (BC). However, in reality, there is typically only one shortest path and the lattice-induced degeneracy could affect the heterogeneity of the BC. We tested this by adding a small random term to the edge lengths to break this degeneracy and better reflect the real-world situation, where a unique shortest path generally exists. We used the following form for the length of an edge $e$: 
\begin{align}
    \ell(e) = \ell_0(e)[1+\delta\ell ((2u-1)]
\end{align}
where $\ell_0(e)$ is the original length of the edge, $\delta\ell=0.1$ and $u$ a random uniform variable in $[0,1]$.  

We compare the results obtained with and without noise and the results are shown in the figure below:
\begin{figure}[h!]
\centering
\includegraphics[width=0.9\linewidth]{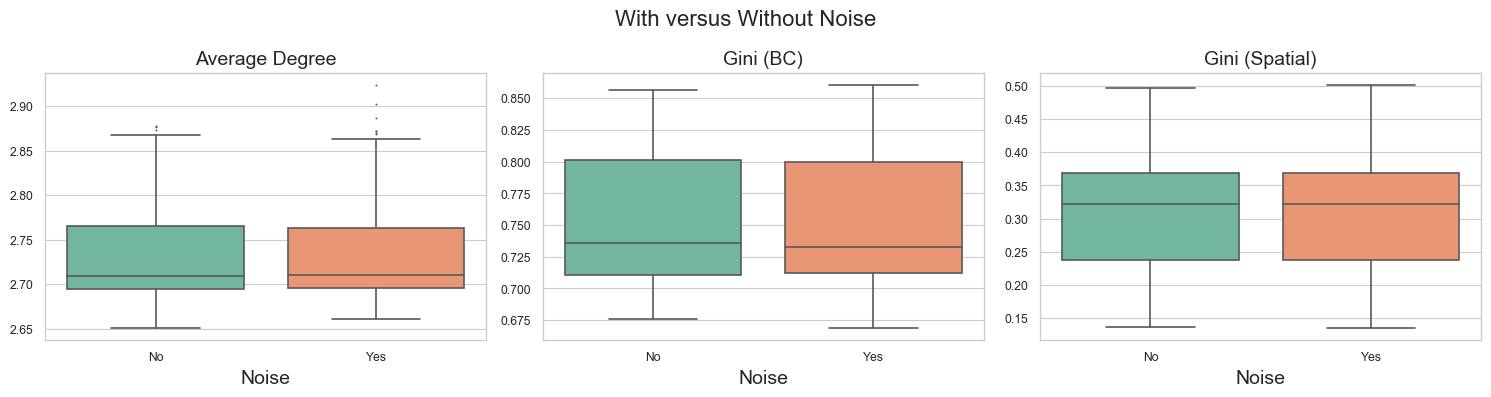}
\caption*{Comparison with and without noise for our different measures (results obtained for $N=2,000$ and for $100$ configurations).}
\label{fig:1}
\end{figure}
We observe that this modification has essentially no effect, with a variation of less than one percent for all measures, including the Gini coefficient of the BC. Its average without noise is $0.7539$, and with noise it is $0.7534$ (representing a change of $-0.08\%$).

\bibliography{bibfile_street_network}